\documentclass[12pt,epsfig]{article}
\usepackage{graphicx}
\usepackage{amsfonts}
\usepackage[mathscr]{eucal}
\usepackage{latexsym}
\usepackage{epsfig}

\def\be{\begin{equation}}
\def\ee{\end{equation}}
\def\ba{\begin{eqnarray}}
\def\ea{\end{eqnarray}}

\begin{document}
\baselineskip=15.5pt
\pagestyle{plain}

\rightline{NUB-3214/Th-01}

\begin{center}

\vskip 1.7 cm

{\large \bf The Mysterious Ultrahigh Energy Cosmic Ray Clustering}\\


\vskip 2.5cm
Luis A. Anchordoqui$^a$, Haim Goldberg$^a$, Stephen Reucroft$^a$, 
Gustavo E. Romero$^{b,c}$,\\ 
John Swain$^a$, and Diego F. Torres$^{b,d}$

\medskip

$^a${\it Department of Physics, Northeastern University\\
Boston, MA 02115, USA}\\
     
\medskip

$^b${\it Instituto Argentino de Radioastronom\'{\i}a\\ 
C.C.5, 1894 Villa Elisa, Buenos Aires, Argentina}

\medskip

$^c${\it Center for Astrophysics, Guangzhou University\\
Guangzhou 510400, China}

\medskip

$^d${\it Department of Physics, Princeton University,\\  NJ 08544,
USA}

\vspace{2cm}

\end{center}

\noindent We examine the correlation between compact radio quasars 
(redshifts in the range $z = 0.3 - 2.2$) and the arrival direction of 
ultrahigh energy cosmic rays forming clusters. Our Monte Carlo simulation 
reveals a statistically significant correlation on the AGASA 
sample: the chance probability of this effect being less than $1\%$. 
The implications of 
this result on the origin of ultrahigh energy cosmic rays are discussed.

\vspace{3.5cm}

\newpage

\section{Introduction}

To date, some 20 giant air showers have been
detected confirming the arrival of cosmic rays (CRs) with nominal energies 
at or above $10^{20} \pm 30\%$ eV, with the record Fly's Eye 
event having $\sim 3.2 \times 10^{20}$ eV \cite{reviews}. The mechanism(s) 
responsible 
for endowing particles with such enormous energies continues to present
a major enigma to high energy physics \cite{bs}. As shown in the pioneering 
works of Greisen-Zatsepin-Kuz'min (GZK) \cite{gzk}, 
the possible sources are constrained by the 
observed particle spectra due to the interaction with the universal radiation 
and magnetic fields on the way to the observer. In particular, any proton 
energy above $5 \times 10^{19}$ eV is degraded by resonant scattering 
via $\gamma +p \rightarrow \Delta \rightarrow p/n + \pi$, such that
less than $20\%$ of protons survive with an energy 
above $3 \times 10^{20}$ ($1 \times 10^{20}$) eV for a distance of 
18 (60) Mpc. A typical nucleus of the cosmic radiation is subject to 
photodisintegration from blue-shifted relic photons, losing about 3-4 
nucleons per traveled Mpc. The mean free path of gamma rays on the radio 
background decreases even more readily. Therefore, if the CR sources are all 
at cosmological distances, the observed spectrum must virtually end
with the GZK cutoff at $E \approx 8 \times 10^{19}$ eV. The spectral cutoff 
is less sharp for nearby sources (within 50 Mpc or so). The arrival directions
of the trans-GZK events are distributed widely over the sky, without 
apparent counterparts (such as sources in the
Galactic Plane or in the Local Supercluster). Moreover, the
data are consistent with an isotropic distribution of sources, in
sharp constrast to the anisotropic distribution of light within
50 Mpc \cite{isotropy}. 

There are two extreme explanations for the observed isotropy. On the one 
hand, it may happen that a bunch 
of sources that by pure chance are 
very close to us dominate the spectrum at the highest energies, and the 
particle orbits are bent \cite{teapot}. This scenario requires large 
scale intervening magnetic fields with intensity ${\cal O}(\mu$G), to 
provide sufficient angular deflection. On the other hand, one can argue 
that there are many 
cosmic ray sources, even at the highest energies.  The lack of plausible 
nearby sources in the arrival direction has encouraged the idea of positing 
undiscovered neutral hadrons, as well as mechanisms 
which are able to break the GZK barrier. Although sufficiently heavy 
particles would avoid the GZK cut off  (the cut-off energy varies as the 
square of the mass of the 
first resonant state) \cite{farrar}, the existence of these particles 
now appears to be 
excluded 
by laboratory experiments \cite{game_over}. The only standard model (SM) 
particle that can reach our galaxy from high 
redshift sources without significance loss of energy is the neutrino. 
The expected event rate for early development of a neutrino shower, however,  
is down from that of an electromagnetic or hadronic 
interaction by six orders of magnitude. This problem can be raised by simply 
postulating that the total neutrino-proton cross section $\sigma_{\nu p}$ 
increases at high center of mass energies, $\sqrt{s}>$ TeV. The hypothesis 
that all 
particles may have a 
strong interaction above collider energies is certainly not new \cite{bere}. 
Recently, some scenarii with $n$ large compact dimensions and precocious 
unification around the TeV-scale have rekindled this idea \cite{ADD}. 
Within this framework, SM 
fields are trapped into a 3+1 dimensional thin shell and only gravity 
propagates in the higher dimensional space. Therefore, the compactification 
radius $r_c$ of the extra dimensions can be large, corresponding to a small 
scale $1/r_c$ of new physics. Here, the weakness of gravitational interactions 
is a consequence of the large compactification radius, encoded in the 
relationship between the Newton constant $G =M^{-2}_{\rm pl}$ and the 
fundamental 
scale of gravity $M_*\sim$ TeV, $M_{\rm pl}^2 \sim r_c^n M_*^{n+2}$.  
From the 4-dimensional perspective, the higher dimensional graviton 
appears as an infinite tower of Kaluza-Klein (KK) excitations \cite{pheno}. 
The weakness of the gravitational interaction can be thus compensated by the 
very large multiplicity of KK states. As a consequence, cross sections 
mediated by spin 2 particles increase rapidly with energy \cite{nu}. 
The KK model may fail \cite{michael} in that the KK modes couple to 
neutral currents, and the scattered neutrino transfers only about 10\% 
of its energy per interaction, thereby elongating the shower profile. 
Nevertheless, an $s^2$ growth of $\sigma_{\nu p}$, supplemented by 
multiple scatters within the nucleus, yields enough energy transfer 
to save the model \cite{jain}. 

On a different track, if some flavor of neutrinos has masses  
$m_{\nu_j}\, {\cal O}(10^{-1})$ eV, the 1.9 K thermal 
neutrino background 
is a target for 
extremely high energy neutrinos to interact forming a $Z$-boson that 
subsequently decays producing 
a ``local'' flux of nucleons and photons \cite{z1}. The energy of the 
neutrino annihilating at the peak of the $Z$-pole is  well above the GZK 
limit
\begin{equation}
E_\nu = \frac{M_Z^2}{2 m_{\nu_j}} = 4 \,\, \left(\frac{{\rm eV}}{m_{\nu_j}}
\right) \times 10^{21}\,\,{\rm eV}\,. 
\end{equation}     
The mean energies of the $\sim $ 2 baryons and $\sim$ 20 $\gamma$-rays 
in each process can be estimated by distributing the resonant energy
among the mean multiplicity of 30 secondaries. The proton energy is given by 
\begin{equation}
<E_p>\, \sim \frac{M_Z^2}{60\, m_{\nu_j}} \sim 1.3 \,\,\left(\frac{{\rm eV}}{m_{\nu_j}}
\right) \times 10^{20} \,\,{\rm eV},
\end{equation}
whereas the $\gamma$-ray energy is given by
\begin{equation}
<E_\gamma>\, \sim  \frac{M_Z^2}{120\,m_{\nu_j}} \sim 0.7 \,\,\left(\frac{{\rm eV}}{m_{\nu_j}}
\right) \times 10^{20} \,\,{\rm eV}.
\end{equation}
The latter is a factor of 2 smaller to account for the photon origin in two 
body $\pi^0$ decay. The annihilation/$Z$-burst rate can be amplified if 
neutrinos are clustered 
rather than distributed uniformly throughout the 
universe \cite{z2}. In such 
a case the probability of neutrinos to annihilate within the 
GZK zone is on the order of 1\%, with the exact value 
depending on unknown aspects of neutrino mixing and relic neutrino clustering 
(more on this below).

Adding to the puzzle, the AGASA experiment has already reported data 
strongly suggesting that the pairing of events on the celestial sky is 
occurring at higher than chance coincidence \cite{agasa}. Specifically, 
four doublets and one triplet of showers with separation angle less 
than the angular resolution $2.5^\circ$ are observed among the 36 events 
reported with mean energy above $4 \times 10^{19}$ eV.  The chance 
probability of observing such a triplet in an isotropic distribution is about
$1\%$. Another doublet is observed if we include events above 
$3.8 \times 10^{19}$ eV. The arrival directions in 
a combined data sample with three other surface experiments further suports 
non-chance association, especially in the direction of the SuperGalactic 
plane \cite{uchihori}. If not a statistical fluctuation, the event clustering
would have profound implications for models discussing the origin of ultrahigh 
energy cosmic rays. In this paper we elaborate on this issue.

\section{Estimates of cluster probabilities}

Let us begin by reviewing the current status of event clustering 
including the recently enlarged sample reported by the AGASA 
experiment \cite{agasa}. To proceed, we adopt the formalism introduced 
in Ref. \cite{haim-tom}. We start 
considering the solid angle $\Omega$ on the celestial sphere covered by an 
experiment divided into $N$ equal angular bins, each with solid angle
$\omega \simeq \pi \theta^2$. 
Then, by tossing $n$ events randomly into
\begin{equation}
N \,\simeq \,\frac{\Omega}{\pi\, \theta^2} = 1045 \,\,\frac{\Omega}
{1 \,\, {\rm sr}} \,\,
\left(\frac{\theta}{1^{^\circ}} \right)^{-2}
\label{N}
\end{equation}
bins, one is left with a random distribution. Now, we identify each event 
distribution by specifying the partition of $n$ total events into a number 
$m_0$ of empty bins, a number $m_1$ of single hits, a number $m_2$ of double 
hits, etc., among the $N$ angular bins that constitute the whole exposure. 
After a bit of algebra, it is easily seen that the probability to obtain 
a given event topology is \cite{haim-tom}
\begin{equation}
P\left(\{m_i\}, n, N\right) = \frac{N!}{N^N}\,\,\frac{n!}{n^n} \,\, \prod_{j=0}
\,\,\frac{(\overline{m_j})^{m_j}}{m_j\,!}\,\,\,,
\label{exact}
\end{equation}
where 
\begin{equation}
\overline{m_j} \equiv N \, \left(\frac{n}{N}\right)^j\, \frac{1}{j\,!}.
\end{equation}
Using Stirling's
approximation for the factorials with the further assumption $N\gg n\gg 1$, 
Eq. (\ref{exact}) can be re-written in a quasi-Poisson form
\begin{equation}
P(\{m_i\},n,N)\approx {\cal P}
\left[ \prod_{j=2} \frac{(\overline{m_j})^{m_j}}{m_j!}
e^{-\overline{m_j}\,r^j (j-2)!} \right]\,,
\label{largeN}
\end{equation}
 where $r\equiv (N-m_0)/n\approx 1$, and the prefactor ${\cal P}$ is given by
\begin{equation}
{\cal P}=e^{-(n-m_1)}\,\left(\frac{n}{m_1}\right)^{m_1 +\frac{1}{2}}\,.
\end{equation}
For ``sparse events'', where $N\gg n$, one expects the number of singlets 
$m_1$ to approximate the number of events $n$. In such a case the prefactor
is near unity.

To estimate the celestial sky coverage $\Omega$
for ground-based experiments,
one rotates the fixed-coordinate solid angle about the earth's axis 
of rotation.
For experiments with vertical acceptance from the zenith to $\theta_z$ the
relevant formula is \cite{haim-tom}
\[
\Omega = 2\pi\left\{
\begin{array} {ll}
2\sin\theta_z\cos\alpha,\quad & {\rm for}\quad \alpha+\theta_z < 90^\circ \\
     1+\sin(\theta_z -\alpha),\quad & {\rm for}\quad \alpha+\theta_z > 90^\circ
\end{array}
\right.
\]
where $\alpha$ is the latitude of the experiment. 

The experimental data reported by Volcano Ranch, Haverah Park, 
Yakutsk and AGASA are used to determine the purely statistical probabilities 
for various cluster topologies. As recommended in 
Ref. \cite{uchihori} only extensive air showers with zenith angle 
$\theta_z <45^\circ$ (that have good quality in the energy and arrival 
direction determination) are taken into account.  The updated data list of 
possible cluster members is quoted in Table 1. 
Using Eqs. (\ref{N}) and (\ref{exact}) we calculate 
the inclusive probabilities for various cluster topologies as a function 
of the angular resolution and the accumulated number of 
events $n$. By inclusive probabilities we mean that the specified 
number of $j$-plets plus any other cluster, counts as all the $j$-plets 
+ extra-clusters. The main experimental properties 
for the four experiments are summarized in Table 2. 

In Fig. 1 we show the 
inclusive probabilities for one and more doublets 
and two and more doublets for the sample of Haverah Park. 
The chance probability for clustering within $3^\circ$ is 
larger than 50\% and hence not statistically significant.
In Fig. 2 we show the inclusive probabilities for 8 doublets and 2 triplets
at different CR-sky coverages. The probability of chance association is
only ``small'' for angular binning tighter than $3^\circ$, and $\Omega > 4$.
The chance probability for clustering within $4^\circ$ and $5^\circ$ remains 
always larger than 10\%. Therefore, the observation of this topology within
the approximate angular resolution of the combined data set, is not 
statistically significant. This result agrees with previous numerical 
simulations \cite{uchihori}.

We now examine
whether there is any evidence for clustering above the statistical
expectation when considering the AGASA subsample. The latter has much better
angular resolution. In Fig. 3 we show the chance probabilities of observing 
5 doublets and one triplet given 58 events at AGASA. The probability is 
extremely sensitive 
to the angular binning. In this case, the chance probability within 
the experimental angular 
resolution is less than $10^{-3}$. This result is in very good agreement with 
the one recently obtained 
using numerical simulations of the angular two point correlation function of 
ultrahigh energy CRs: A $3 \times 10^{-4}$ probability of chance clustering 
with a bin size of $2.5^\circ$ and an energy cut-off at 
$4.8 \times 10^{19}$ eV \cite{TT1}. Furthermore, when including data from 
Yakutsk's experiment above $2.4 \times 10^{19}$ eV, the combined probability 
of chance clustering is reported to be as small as $4 \times 10^{-6}$,
strongly suggesting that CR sources are point-like on cosmological 
scales \cite{TT1}.

\section{Correlation with high redshift objects} 

Compact radio quasars (CRQSOs) are strong radio emitters, a fact that 
along with their variability, is indicative of strong beaming. The bulk 
of the observed non-thermal emission of these objects is thought to be 
produced in strong, relativistic jets of charged particles emitted by 
the active nucleus, which is likely formed by an accreting supermassive 
black hole. These powerful objects have been under suspicion as the 
primary source of ultrahigh energy CRs for some time 
now \cite{bf,sigletal,vir}. Therefore, we find it particularly attractive 
to examine whether there exists
a correlation between CR-clusters and CRQSOs. We shall use the 451 CRQSOs 
with flat spectrum and declination above $-10^\circ$ degrees taken from 
the surveys of Ref. \cite{kuhr}. With the 
aim of finding the positional coincidences and evaluating their significance, 
we adopt the procedure 
of Ref. \cite{sigletal}. First, we look for real correlations between the 
two sets. In order to do so, we consider a circle around the centroid of 
each CR event, this circle has a radius equal to the reported 1 sigma
positional error (see Table 2). If a CRQSO is within the circle of all 
members of the cluster, we say that there is a positional coincidence.
We are not giving a higher significance to directional coincidences with
small offsets than to coincidences that are not so close, just
because the original errors of the CRs are of the order of
degrees. As a first trial, we look for positional coincidences between the 
CRQSO sample and all the events listed in Table 1.
We have found that there are no objects which correlate with the 
direction of triplets, so in the following a cluster denotes a pair of events.
Just 4 clusters (out of 10) of the sample are positionally coincident with 
CRQSOs.
Numerical simulations using large numbers of synthetic populations
(thousands of them were made for each correlation study) sampled
randomly and uniformly in right ascension and declination, are
then performed in order to determine the probability of pure
chance spatial association. Strictly speaking, we generate
synthetic populations of 451 CRQSOs and 
compare them with the actual positions of the CRs. We have taken into
account that the artificial sets of CRQSOs are constrained
(as are the actual ones) to the declination range $\delta
> -10^\circ$. In Fig. 4 we show the sky distribution of 451 CRQSOs in 
equatorial
coordinates. The apparent anisotropy is due to obscuration from the galactic
plane of our galaxy. Since only one ultrahigh energy CR cluster lies in 
this region, the
real level of positional coincidences (even considering that an uniform
distribution of CRQSOs is there) cannot significantly differ from what is
observed. The present Monte Carlo simulations preserve an isotropic
distribution for background sources. The level of random positional coincidence
after 5000 simulations (a larger number of simulations 
does not significantly modify the result) is shown in Fig. 5. Notice that 
the actual 
coincidences are less than 1 standard deviation away from the simulated 
mean value $3.5 \pm 1.8$. As a second trial, we repeat the whole analysis 
but considering only CRs in the AGASA sample above $4 \times 10^{19}$ eV. 
In this case we found that the number of real matches is 3, whereas the 
expected number from pure chance estimated from the simulations is $0.45 \pm 
0.66$, i.e., $3.8$ standard deviations below (see Fig. 6). The Poisson 
probability 
of a random occurrence of any number of coincidences greater than or 
equal to the 
real positional coincidence is $P = 9.7 \times 10^{-3}$.  

\section{Outlook}

Let us end with a discussion (and some speculations) on the 
implications of our result. The clustering beyond statistics by itself 
imposes certain 
constraints on possible CR-sources. As can be read in Table 1, the 
event-pairing has members with rather different energies. 
The energy spread would have profound consequences for the 
propagation of charged ultrahigh energy CRs, even in the regular 
magnetic field of ${\cal O}$(nG). Strictly speaking, if the Larmor radius 
of a particle ($r_{\rm L} \simeq 10^2$ Mpc $E_{20}/B_{-9}$) is much larger 
than the coherence length of the magnetic field $\ell_{\rm coh}$, the typical 
deflection angle from the direction of the source, located at a distance $d$, 
can be estimated assuming that the particle makes a random walk in the 
magnetic field \cite{we}
\begin{equation}
\theta(E) \simeq 3.8^\circ\, \left(\frac{d}{50\,\,{\rm Mpc}}\right)^{1/2}\,\,
\left(\frac{\ell_{\rm coh}}{1 \,\, {\rm Mpc}} \right)^{1/2} \,\, 
\left(\frac{B_{-9}}{E_{20}} \right)\,\,,
\end{equation}
where $E_{20}$ is the energy of the particle in units of $10^{20}$ eV, and 
$B_{-9}$ is the magnetic field in units of $10^{-9}$ G. Now,
it is straightforward to check that scatterings 
in large scale magnetic irregularities ${\cal O}$ (nG) \cite{b} 
are enough to bend the orbits of trans-GZK protons more than $5^\circ$ in a 
50 Mpc trip. The time delay in the arrival of cosmic ray pairs can be used 
to impose
additional constraints on the nature of the sources. In particular, the
time lags shown in Table 1 are long enough to rule out an origin based on
bursting sources. The average time delay in the pairs of the sample 
(with real positional coincidences) is 3 yr
$\pm$7 months, favoring a compact (but not transient) source. Typical
source sizes for the production of high energy particles in AGNs are
smaller that the radius of the outer gamma-spheres, i.e. $<10^{15}$ cm
\cite{BL}.
All in all, if forthcoming data confirm that the clustering is not a
statistical artifact, we are left with two main possibilities: (i) the
particles are charged but then the sources should be very nearby
(which is hard to accept already), or (ii) 
the CR constituents of the clusters are neutral particles! 
The first option is very problematic
because if the sources are nearby (say located within our own Galaxy)
global isotropy should not be observed, contrary to the evidence, unless
strong magentic fields could produce the required isotropization, in which
case the clustering effect is broken.                     

In order to explain the event-pairs (pointing towards distant sources) 
by neutrino showers with TeV-scale quantum gravity, neutrinos must be 
capable of generating vertical atmospheric cascade developments both at 
the lower and higher energy pair. The rapid energy behaviour of cross 
section in KK models makes this requirement rather difficult \cite{dom}. 
The $Z$-burst, 
however, still appears as a viable model. As stated above the secondaries 
produced by $Z$ decay are mainly photons. Because cascading by $e^+e^-$ pairs 
(produced by double pair 
production on the microwave background, or pair production on the radio 
background) is rapid, the photon energy is quickly reduced below 
the GZK-limit. Just to get an estimate, 
the mean interaction length at $3 \times 10^{20}$ eV is a 
few Mpc, and the 
energy attenuation length (which is very sensitive to the magnetic field 
because of synchrotron losses) assuming extragalactic magnetic 
fields ${\cal O}$ (nG) is $\sim 20$ Mpc. Both  increase with 
decreasing energy. As mentioned above, some protons 
are able to survive a 60 Mpc trip with energies above $10^{20}$ eV. 
Besides, the nucleon-channel would suffer magnetic field deflections 
reducing the correlation with the neutrino-emitting-source. For instance,
$B_{-9} \sim 2$ would defocus a proton $E_p \sim 1.5 \times 10^{20}$ eV 
(as the one detected by AGASA on 97/03/30) more than $5^{\circ}$ 
\cite{sigletal}. Therefore, since the photon flux is depleted through 
interactions with the radiation backgrounds, the relation 10:1 of 
photon/nucleon in the $Z$ decay is significantly reduced, reducing also 
the correlation with CRQSOs (because the protons are bent in the magnetic 
field). In other words, the $Z$-burst model predicts a 
correlation between ultrahigh energy CRs and  highly redshifted 
background sources \cite{vir}, which 
diminishes with increasing energy (say $E> 8 \times 10^{19}$ at 1 $\sigma$ 
level) \cite{sigletal}. We note that none of the members of clusters 
with real matches in our analysis satisfy the above energy cut-off. 
We do not attempt to make yet another estimate here and only note that on  
average, proton deflections of $\simeq 10^\circ$ \cite{TT2} are 
consistent with existing bounds on neutrino masses, relic neutrino density 
and magnetic field strengths. 

In closing we would like to note that the $Z$-burst model is not free of 
problems: Firstly, it is desirable to understand more completely the 
depletion of high energy photon flux, since the characteristcs of the 
air showers detected so far do 
not seem to resemble an electromagnetic $\gamma$-ray shower \cite{g-showers}. 
Secondly, there is the additional difficulty of getting neutrinos with 
energies $\sim 10^{22}$ eV. If these are secondaries from pion production, 
this implies that the primary protons which produce them must have energies 
of $\sim 10^{23}$ eV. While standard acceleration mechanisms require quite 
extreme parameters to achieve such extraordinary energy, we note that 
only a few dozen of such sources in the whole visible Universe would suffice.

{\it Mysterious: adj. Having an import not 
apparent to the senses nor obvious 
to the intelligence; beyond ordinary understanding} \cite{mys}. This appears 
to be the situation concerning the seeming ultrahigh energy cosmic ray 
clustering discussed in this paper.

\section*{Acknowledgments}

We would like to thank Gabor Domokos and Susan Kovesi-Domokos for valuable 
discussions. We are also grateful to Martin Pessah and Paula Turco for their 
help in the make up of Fig. 4.
This work has been partially supported by CONICET (LAA, GER, DFT), the 
National Science Foundation (HG,SR,JS),  Fundaci\'on 
Antorchas (GER, DFT) and ANPCT (GER). GER is very grateful to Prof. 
Junhui Fan for his kind hospitality during his stay at the Center for 
Astrophysics, Guangzhou University, where part of this work was done.

\newpage
\begin{figure}
\label{c1}
\begin{center}
\epsfig{file=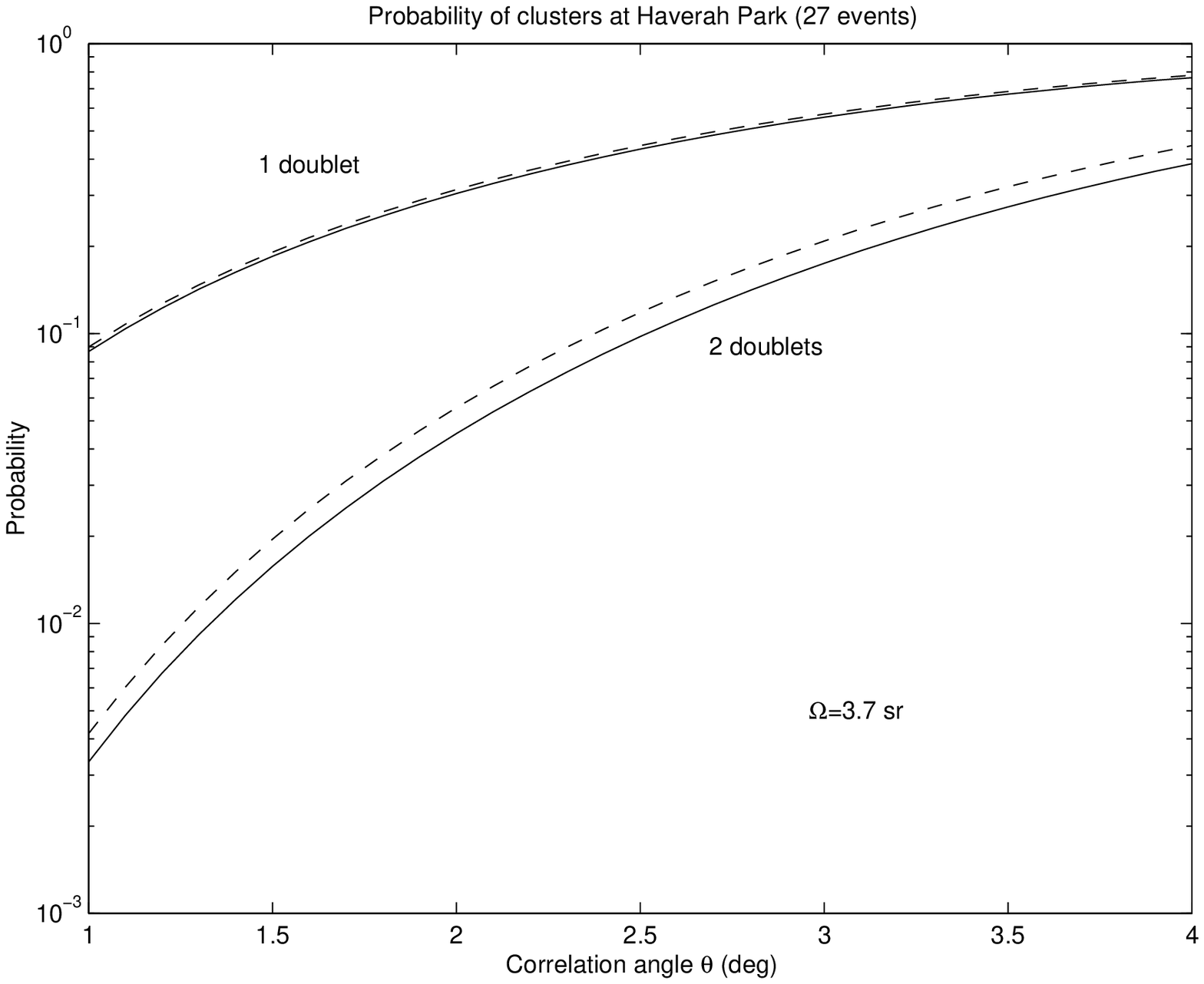,width=15.cm,clip=} 
\caption{Inclusive probabilities for various clusters in a 27 event sample 
as seen by the Haverah Park experiment. 
The solid line is the exact result, 
whereas the 
dashed line is the Poisson approximation.}
\end{center}
\end{figure}

\newpage
\begin{figure}
\label{c2}
\begin{center}
\epsfig{file=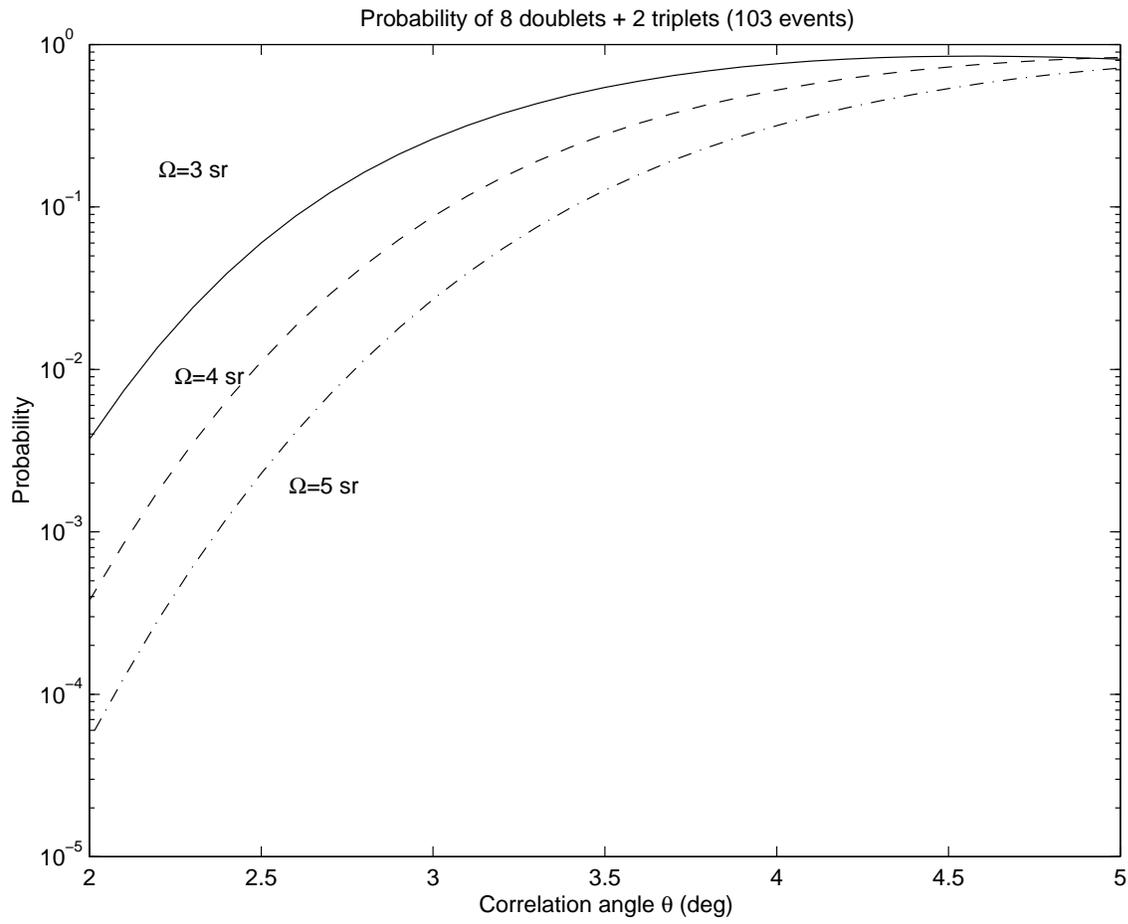,width=15.cm,clip=} 
\caption{Inclusive probabilities for 8 doublets and 2 triplets 
in a 103 event sample for various celestial solid angles.}
\end{center}
\end{figure}

\newpage
\begin{figure}
\label{c3}
\begin{center}
\epsfig{file=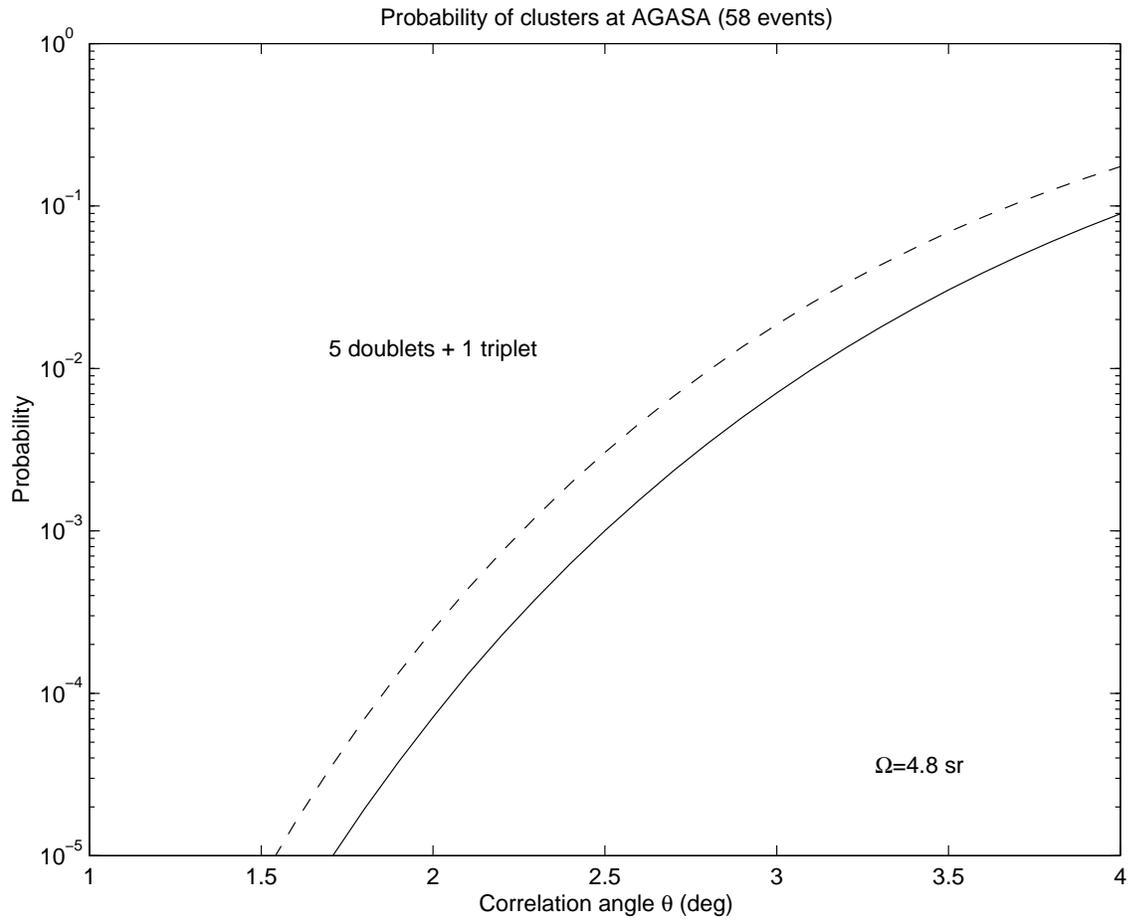,width=15.cm,clip=} 
\caption{Inclusive probabilities for 5 doublets and 1 triplet in a 58 event sample at AGASA. Solid (exact), dashed (Poisson).}
\end{center}
\end{figure}

\newpage
\begin{figure}
\label{c4}
\begin{center}
\epsfig{file=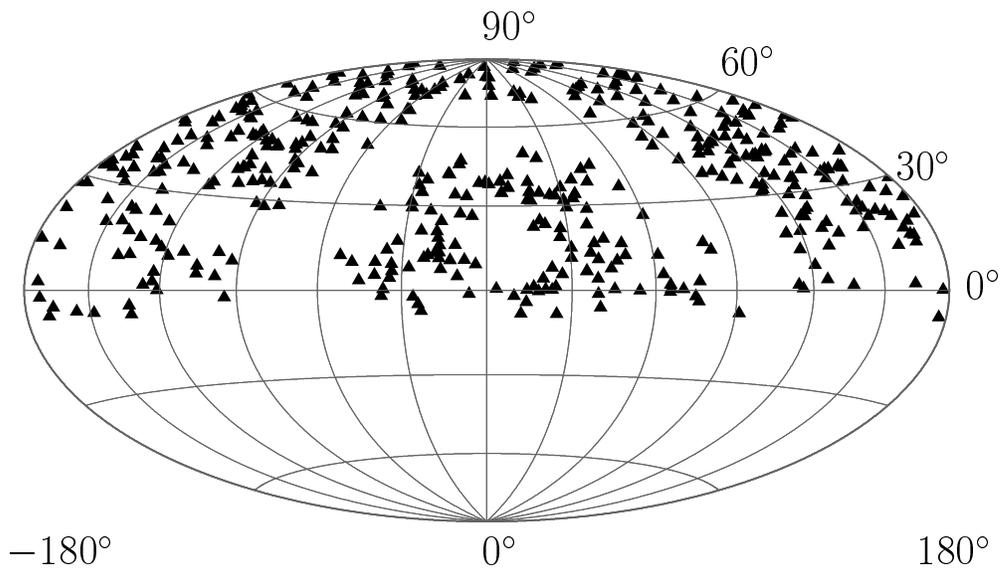,width=14.cm,clip=} 
\caption{Sky distribution of 451 CRQSOs in equatorial coordinates. The 
sample is
complete within the sensitivity of the radio surveys (see Ref. \cite{kuhr}). 
The
apparent anisotropy is due to obscuration from the galactic plane of our
galaxy.}
\end{center}
\end{figure}

\newpage

\begin{figure}
\label{c5}
\begin{center}
\epsfig{file=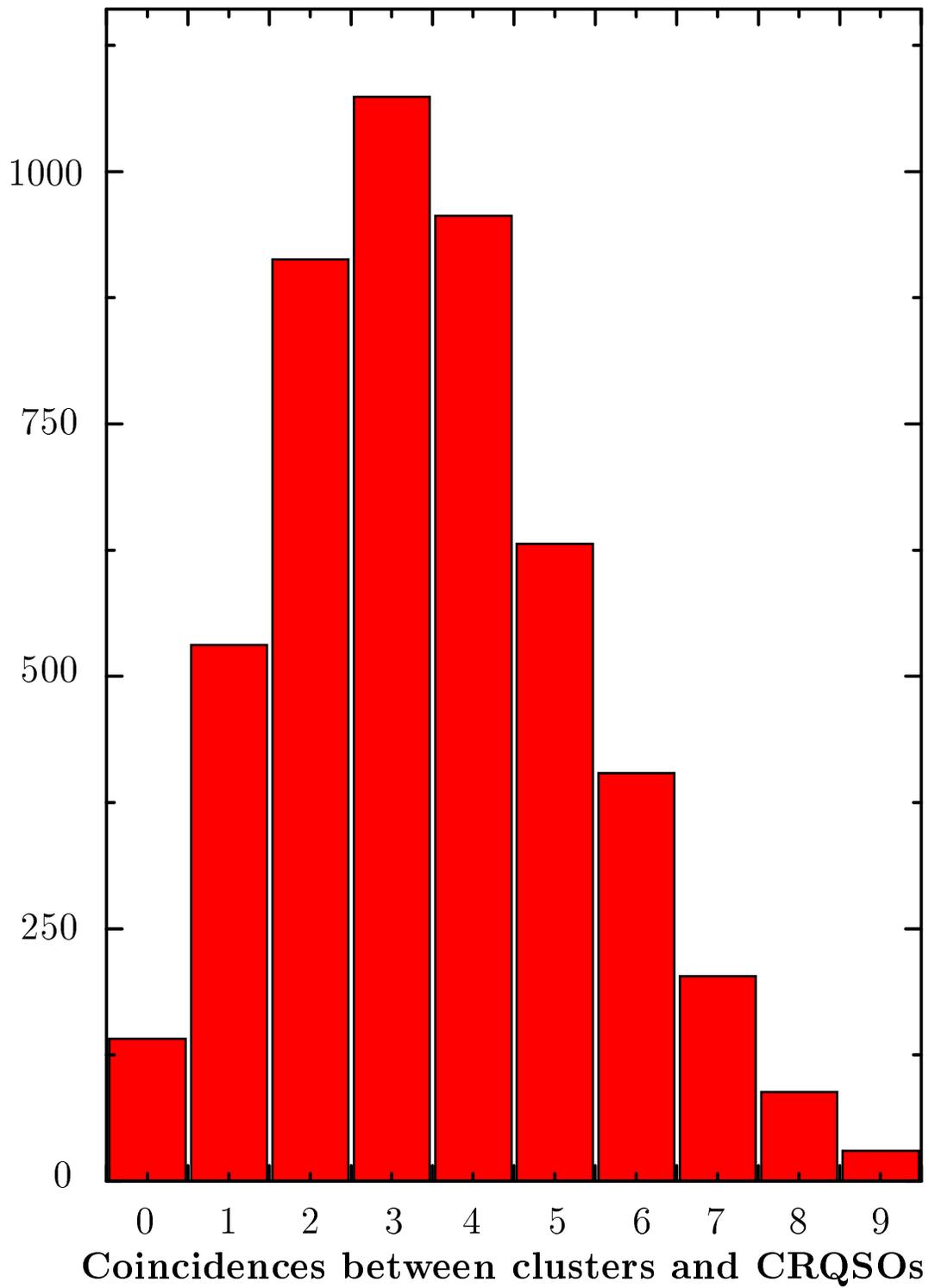,width=15.cm,clip=} 
\caption{Simulated positional coincidence between  
all CR-clusters taken from Table 1 and CRQSOs for 5000 runs.}
\end{center}
\end{figure}

\newpage
\begin{figure}
\label{c6}
\begin{center}
\epsfig{file=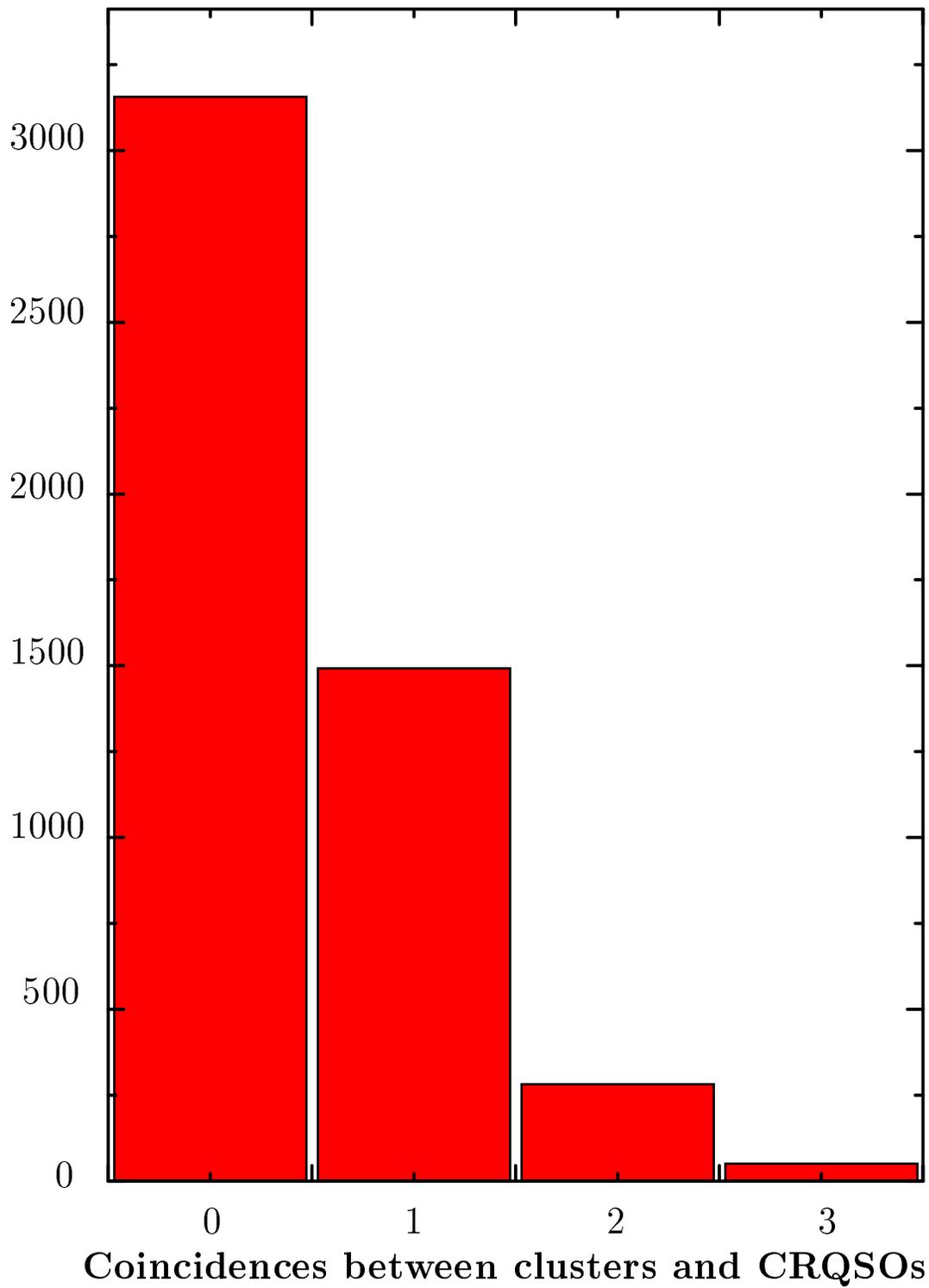,width=15.cm,clip=} 
\caption{Simulated positional coincidence between  
AGASA-clusters with energies above $4 \times 10^{19}$ eV and CRQSOs for 5000
runs.}
\end{center}
\end{figure}

\newpage

\begin{table}
\begin{center}
\begin{tabular}{|r||r||r||r||r|r|} \hline
 Cluster & Experiment & Date & Log E & R.A. & Dec. \\ \hline \hline
 Triplet \#1 & Haverah Park & 810105 & 19.99 & 20.00 & 20.00 \\
             & AGASA & 931203 & 20.33 & 18.91 & 21.07 \\
              & AGASA & 951029 & 19.71 & 18.53 & 20.03 \\ \hline
 Triplet \#2  & AGASA & 920801 & 19.74 & 172.30 & 57.14 \\
              & AGASA & 950126 & 19.89 & 168.65 & 57.58  \\
              & AGASA & 980404 & 19.73 & 168.44 & 55.99 \\ \hline  \hline
 Doublet \#1  & AGASA & 910420 & 19.64 & 284.90 & 47.79  \\
              & AGASA & 940706 & 20.03 & 281.36 & 48.32 \\ \hline
  Doublet \#2  & AGASA & 860105 & 19.74 &  69.03 & 30.15 \\
              & AGASA & 951115 & 19.69 &  70.39 & 29.85 \\ \hline
Doublet \#3 & Haverah Park &  860315 & 19.71 & 267.00 & 77.00 \\
         & AGASA & 960513 & 19.68 & 269.05 & 74.12 \\ \hline
   Doublet \#4  & Haverah Park & 720525 & 19.65 & 239.00 & 79.00 \\
              & Yakutsk & 911201 & 19.62 & 235.40 & 79.80 \\ \hline
 Doublet \#5  & Volcano Ranch & 610319 & 19.73 & 154.10 & 66.70 \\
              & Haverah Park & 850313 & 19.62 & 157.00 & 65.00 \\ \hline
 Doublet \#6  & Haverah Park & 661008 & 19.67 & 164.00 & 50.00 \\
              & Yakutsk & 750317 & 19.67 & 163.70 & 52.90 \\ \hline
 Doublet \#7  & Haverah Park & 740228 & 19.86 & 264.00 & 58.00 \\
              & AGASA & 980330 & 19.84 & 259.16 & 56.32 \\ \hline
 Doublet \#8  & Haverah Park & 760206 & 19.62 & 165.00 & 64.00 \\
              & Haverah Park & 850313 & 19.62 & 157.00 & 65.00 \\ \hline
 Doublet \#9  & AGASA & 960111   & 20.16  & 241.5  & 23.00 \\ 
              & AGASA & 970410  & 19.58  & 239.50 & 23.70 \\ \hline
 Doublet \#10 & AGASA & 961224 & 19.70 & 213.75 & 37.7 \\ 
              & AGASA & 000526 & 19.70 & 212.00 & 37.1 \\ \hline
\end{tabular}
\end{center}
\caption{\it Updated list of triplets and doublets within space angles of
$3^\circ$, $4^\circ$ and $5^\circ$. }
\label{tab:events}
\end{table}

\begin{table}
\begin{center}
\begin{tabular}[b]{|c||c|c|c|c|c|c|c|} \hline
  Experiment & Begin & End/status & Latitude & Longitude & $\Omega$ (sr) & $\theta_{\rm min}$ & $n$  \\
\hline \hline
  AGASA & 1990 & in operation & $35^\circ \, 47'$ N & $138^\circ \,30'$ E 
& 4.8 & 1.8$^\circ$ & 58 \\ \hline
  Haverah Park & 1968 & 1987 & $53^\circ\, 58'$ N 
& $1^\circ\, 38'$ W & 3.7 & $3.0^{\circ}$ & 27  \\ \hline
  Yakutsk & 1972 & in operation & $61^\circ \, 42'$ N & $129^{\circ}\, 24'$ E 
& 3.1 &  $3.0^{\circ}$& 12  \\ \hline
  Volcano Ranch & 1959  & 1963 & $35^{\circ} \,09'$ N & 
$106^\circ\, 47'$ W & 5.0 &  $3.0^{\circ}$  & 6 \\
  & 1972 & 1974 & & & & & \\ \hline 
\end{tabular}
\end{center}
\caption{\it Event rates, celestial solid angles 
$\Omega$ (sr), and angular resolutions $\theta_{\rm min}$.
We remark that the celestial solid angles have been reduced by a factor 
of 0.7 (see \cite{haim-tom} for details).}
\end{table}


\begin{thebibliography}{99}
\bibitem{dom} Note that neutrino showering below the GZK-energy cannot 
be rulled out within the string theory framework. 
See G. Domokos and S. Kovesi-Domokos in Ref. \cite{nu}.
\end{thebibliography}
\end{document}